\documentclass[12pt]{article}
\usepackage{amssymb}
\usepackage{graphicx}
\begin{document}
\title{Quantum incompressibility and Razumov Stroganov type conjectures}
\author {  Vincent Pasquier.}
\date{}
\maketitle \hskip-6mm { Service de Physique Th\'eorique, C.E.A/
Saclay, 91191 Gif-sur-Yvette, France.}

\begin{abstract}
We establish a correspondence between polynomial representations of
the Temperley and Lieb algebra and certain deformations of the
Quantum Hall Effect wave functions. When the deformation parameter
is a third root of unity, the representation degenerates and the
wave functions coincide with the domain wall boundary condition
partition function appearing in the conjecture of A.V. Razumov and
Y.G. Stroganov. In particular, this gives a  proof of the
identification of the sum of the entries of a $O(n)$ transfer matrix
eigenvector and a six vertex-model partition function, alternative
to that of P. Di Francesco and P. Zinn-Justin.

\end{abstract}
\maketitle

\section{Introduction}

This paper is aimed at establishing a correspondence between the
deformation of certain wave functions of the Hall effect and
polynomial representations of the Temperley and Lieb (T.L.) algebra.

This work originates from an attempt to understand the conjecture of
A.V. Razumov and Y.G. Stroganov
\cite{razumov1}\cite{razumov2}\cite{pierce}, and some partial
results towards its proof by P.Di Francesco and P. Zinn-Justin
\cite{Pdf}.

We consider the analogue of spin singlet wave functions of the Hall
effect when one deforms the permutations into the braid group. This
amounts to analyze some simple representations of the T.L. algebra
on a space of polynomials in $N_e$ variables where $N_e$ is the
number of electrons. The relation with the Hall effect arises when
we require certain incompressibility properties.

One of the wave functions we consider here is the Halperin wave
function \cite{halperin} for a system of spin one half electrons at
filling factor two. When the deformation parameter $q$ is a third
root of unity, the braid group representation degenerates into a
trivial representation. In this way, we obtain a proof alternative
to, and apparently simpler than, that given in \cite{Pdf} of the
equality between the sum of the components of the transfer matrix
eigenvector and the six vertex model partition function with domain
wall boundary conditions \cite{razumov2}\cite{korepin}.

Another wave function we consider is the Haldane Rezayi wave
function \cite{haldanerezai}\footnote{More precisely a minor
modification of it considered in \cite{hallpasq}.} describing a
system of electrons of spin one half at filling factor one. This
wave function is a permanent, and its deformation is described in
terms of Gaudin's determinants \cite{Gaudin}. When $q$ is a third
root of unity, it degenerates to the square of the six vertex model
partition function.

In a separate publication \cite{pas2}, we shall consider the Moore
Read wave function describing spinless bosons at filling factor one
\cite{mooreread}. Its deformation involves an extension of the braid
group known as the Birman-Wenzl algebra \cite{birman} which can be
represented on a polynomial space similarly to the cases presented
here. In some appropriate limit, the representation degenerates and
the wave function coincides with the transfer matrix eigenvector
considered in \cite{Pdf2} related to the conjecture of J.De Gier and
B. Nienhuis \cite{degier}.

In general, when a Quantum Hall Effect wave function is discovered,
it is soon after observed experimentally. We argue here, that as a
bonus, Quantum Hall Effect wave functions and their deformations
yield nice mathematical objects. Moreover, all these objects seem to
be in relation with striking conjectures emanating from the six
vertex model.

\bigskip

Since the permutation group relevant in the quantum Hall effect is
technically simpler than the braid group case, let us for
pedagogical reasons explain why finding a wave function turns out to
be a useful tool to obtain a polynomial representations of the
permutation algebra. Essentially, the rest of the paper extends the
idea presented here to the braid group case.

We consider electrons in a strong magnetic field projected in the
lowest Landau level. In a specific gauge the orbital wave functions
are given by:
\begin{eqnarray}
 \psi_{n}(z)= {z^n\over \sqrt{n!} } e^{- {z \bar z \over 4 l^2}},\label{orbitals}
\end{eqnarray}
where $z=x+iy$ is the coordinate of the electron, and $l$ the
magnetic length defines the length scale related to the strength of
the magnetic field. These orbitals are shells of radius $\sqrt
{2n}l$ occupying an area $2\pi l^2$. Each orbital $n$ is represented
by a monomial $z^n$.

The quantum Hall effect \cite{prange} ground state $\Psi$ is
obtained by combining these individual orbitals into a manybody wave
function. A monomial $z_1^{\lambda_1}...z_{N_e}^{\lambda_{N_e}}$
describes a configuration where the electron $j$ occupies the
orbital $\lambda_j$. The wave function is a linear combinations of
such monomials. The effect of the interactions is to impose some
vanishing properties when electrons are in contact: $\Psi \sim
(z_i-z_j)^m$ with $m$ an integer when $z_i-z_j\to 0$.

The physical properties are mainly characterized by the filling
factor $\nu$ which is the number of electrons per unit cell of area
$2\pi l^2$. When the filling factor is equal to $\nu$, the
accessible orbitals and thus the maximal degree in each variable is
bounded by  $\nu^{-1} N_e$. On the other hand, the effect of the
interactions ($m$) is to force the electrons to occupy more space,
thus to occupy higher orbitals and and has the effect of increasing
the degree. The problem is thus to obtain wave functions with the
maximal possible filling factor (equivalently the lowest degree in
each variable) compatible with the vanishing properties imposed by
the interactions.

Once such a wave function is obtained, it is the nondegenerate
lowest energy state of a Hamiltonian invariant under the
permutations, thus we know that it is left invariant under the
permutations. By disentangling the coordinate  part from the spin
part, we obtain an irreducible representation of the permutation
algebra acting on polynomials.

Let us illustrate this point in the case of the Halperin wave
function \cite{halperin} which describes a system of spin one half
electrons at filling factor two. There are no interactions between
the electrons, but due to the Pauli principle, the wave function
must vanish when two electrons of the same spin come into contact.
Each independent orbital can be occupied with two electrons of
opposite spin, which is why the maximal filling factor is equal to
two.

An equivalent way to impose the constraint is to require that any
linear combination of the spin components of the wave function
vanishes when three electrons come into contact. The reason for this
is that two of the electrons involved will necessary have the same
spin. When this constraint is taken into account with the minimal
degree hypotheses, one obtains a space of polynomials which can be
recombined with the spin components into a wave function changing
sign under the permutations. Thus we know a priori that the spatial
part of the wave function carries an irreducible representation of
the permutation algebra dual to that of the spins. This is precisely
by generalizing this argument to the braid group case that we obtain
the representations of the T.L. algebra mentioned above.

In the  permutation group case case, the components have the simple
structure of a product of two Slater determinants grouping together
the electrons with the same spin and one does not require to
recourse to this machinery.

Let us now briefly indicate why the Halperin wave function may have
something to do with the eigenvector of a transfer matrix in the
link pattern formulation \cite{batchelor}. The wave function is a
spin singlet, and the spin components can best be described in a
resonating valance bond (RVB) picture as follows: The labels of the
electrons are disposed cyclically around a circle and are connected
by a link when two electrons form a spin singlet. Links are not
allowed to cross in order to avoid overcounting states. These RVB
states coincide with the link patterns of \cite{batchelor}. Thus,
the Halperin wave function as the eigenvector of the transfer matrix
develops on a basis of link patterns. By deforming the permutation
action on link patterns into a T.L algebra action, one is forced to
deform accordingly the polynomial representation so as to insure the
invariance of the total wave function. When $q$ is a third root of
unity, this property is shared by the transfer matrix eigenvector
and allows to identify the two.

In the braid group case, the situation is technically more involved
than for the permutations. Nevertheless, the minimal degree
hypothesis combined with some annulation constraint satisfied by
linear combination of the spin components yields a wave function
with the correct invariance properties. A major difference with the
Hall effect is that the cancelation no longer occurs at coincident
points, but at points shifted proportionally to the deformation
parameter $q$. Typically, we require that for three arbitrary
electron labels $i<j<k$ ordered cyclically, the wave function
vanishes when the corresponding coordinates take the values $z,\
q^2z, q^4z$.

 One is also led to study the
affine extension in order to impose cyclic invariance properties
which are tautologically satisfied with the permutations. While
defined in a natural way on the link patterns, the cyclic properties
require to introduce a shift parameter $s$ when we identify the
coordinate $i+N_e$ with the coordinate $i$: $z_{i+N_e}=sz_i$. When
this shift parameter is related in a specific way to the braid group
deformation parameter, the generalized statistics properties can be
established coherently. Here, $z_{i+N_e}=q^6 z_i,$ but the same
annulation property can also be satisfied with $s$ not related to
$q$, and this can be achieved at the price of doubling the degree
and enlarging the algebra \cite{pas2}.

In the  the Haldane Rezayi case, \cite{haldanerezai}, the
interactions are such that the wave function must vanish as the
square of the distance when electrons of the same spin come into
contact. For the same reason as before, this amounts to impose that
any linear combination of its spin components vanishes as the square
of the distance when three electrons come into contact. This wave
function is a permanent, and its deformation is described in terms
of Gaudin's determinants \cite{Gaudin}. It degenerates to the square
of the six vertex model partition function when the deformation
parameter is a third root of unity.

The paper is organized as follows. In section \ref{hecke-algebra},
we recall some properties about Hecke algebras and their polynomial
representations. Section \ref{T.L. algebra} introduces the T.L.
algebra representation used here. Section \ref{Q.H.E.deformation} is
the core of the paper where we work out the deformed Hall effect
wave functions.

We have attempted to be self contained, but in order not to overload
the text with technicalities, we have relegated most of the proofs
to appendices to which we refer when it is useful.

\section{Hecke Algebra.\label{hecke-algebra}}

In this section, we recall some known facts about the Hecke and
Temperley and Lieb algebras \cite{Pasthese}\cite{Jones}.

The Braid group algebra is generated by the braid group generators
$t_{1},t_{2},...,t_{n-1}$, obeying the braid relations:
\begin{eqnarray}
 t_{i}t_{i+1}t_{i}&=& t_{i+1}t_{i}t_{i+1}\cr
 t_{i}t_{j}&=&t_{j}t_i\ \
{\rm if} \ |i-j|>1, \label{braid}
\end{eqnarray}
for $1\le i\le n-1$. It can be convenient to use the notation
$t_{ii+1}$ instead of $t_{i}$, and we will use it when necessary.
The Hecke algebra is the quotient of the Braid group algebra by the
relations:
\begin{eqnarray}
(t_{i}-q)(t_{i}+{1\over q})=0,\label{hecke}
\end{eqnarray}
It can also be defined using the projectors $e_{i}=t_{i}-q$ obeying
the relations:
\begin{eqnarray}
e_{i}^2&=&\tau e_{i}, \cr e_ie_j&=&e_je_i\ \ {\rm if} \ |i-j|>1\cr
e_{i}e_{i+1}e_{i}-e_{i}&=&
e_{i+1}e_{i}e_{i+1}-e_{i+1}.\label{hecke1}
\end{eqnarray}
where we set $\tau=-(q+q^{-1})$. The Temperley-Lieb (T.L.) algebra
$\mathcal{ A}_n$ is the quotient of Hecke algebra by the relations:
\begin{eqnarray}
e_{i}e_{i+1}e_{i}-e_{i}= e_{i+1}e_{i}e_{i+1}-e_{i+1}=0.\label{T.L.}
\end{eqnarray}
In $\mathcal{ A}_n$, a trace can be defined \cite{Jones} as:
\begin{eqnarray}
 {\rm t r}(xe_p)=\tau^{-1} {\rm t r}(x) \ \forall x\in \mathcal{
 A}_p.
 \label{trace}
\end{eqnarray}

The affine Hecke algebra, \cite{BGHP}\cite{Macdo2}\cite{lecture}, is
an extension of the Hecke algebra (\ref{hecke}) by generators $y_i,\
1\le i\le n$ obeying the following relations:
\begin{eqnarray}
&a)&\ \ y_iy_j=y_jy_i\cr &b)&\ \ t_iy_j=y_jt_i \ \ \ {\rm if}\ j\ne
i,i+1 \cr &c)&\ \ t_{i}y_{i+1}=y_{i}t_{i}^{-1}\ \ {\rm if} \ i\le
n-1.\label{affine-gener}
\end{eqnarray}
In (\ref{discussion}), we indicate why (\ref{affine-gener}c) is
natural from the Yang-Baxter algebra point of view.

This algebra can be endowed with two possible involutions:
$e_i^*=e_i,\ y_i^*=y_i^{\pm 1}$, $q^*=q^{\pm 1}$.

The symmetric polynomials in the $y_i$ are central elements.

We define the affine T.L. algebra $\mathcal{ A}_n'$ as the extension
of the T.L. algebra (\ref{T.L.}) by the generators $y_i$.

\subsection{Yang's realization of the Affine relations.}

The commutation relations of the affine generators $y_i$ become
simpler to understand if we assume that we have a representation of
the permutations $k_{ij}$ acting in the natural way on the indices.
Let us introduce the operators $x_{ij}=t_{ij}k_{ij}$ for $i<j$ and
$x_{ji}=x_{ij}^{-1}$. These operators obey the Yang's relations:
\begin{eqnarray}
x_{ij}x_{ji}&=&1,\cr  x_{ij}x_{kl}&=&x_{kl}x_{ij}\ \forall i\ne j\ne
k\ne l,\cr x_{ij}x_{ik}x_{jk}&=&x_{ik}x_{jk}x_{ij}. \label{yang-op}
\end{eqnarray}
We also assume that we have commuting operators $s_i$ such that
$s_is_jx_{ij}=x_{ij}s_is_j$.

Using (\ref{yang-op}), one verifies that the operators  defined as
\cite{BGHP}\cite{lecture}\cite{affine}:
\begin{eqnarray}
y_1&=&x_{12}x_{13}...x_{1n}s_1\cr
y_2&=&x_{23}x_{24}...x_{2n}s_2x_{21}\cr y_n&=&s_n
x_{n1}x_{n2}...x_{nn-1} . \label{yang-rep0}
\end{eqnarray}
commute. Indeed, they coincide with the scattering matrices of Yang
\cite{Yang}. (\ref{affine-gener}b) follows directly from
(\ref{yang-op}) once we substitute $t_i=x_{ii+1}k_{ii+1}$.
(\ref{affine-gener}c) is a direct consequence of the definition
(\ref{yang-rep0}) of $y_i$.

Gathering the permutation operators $k_{ij}$ together, we can obtain
another presentation of the $y_i$. Let us introduce the cyclic
operator:
\begin{eqnarray}
\sigma=k_{n-1n}...k_{23}k_{12}s_1.\label{sigma-k}
\end{eqnarray}
Then we have:
\begin{eqnarray}
y_1&=&t_{1}t_{2}...t_{n-1}\sigma \cr
y_2&=&t_{1}^{-1}y_1t_{1}^{-1}\cr
y_n&=&t_{n-1}^{-1}y_{n-1}t_{n-1}^{-1}.\label{affine-rep1}
\end{eqnarray}

We can define an additional generator to the $t_{i}$: $t_{n}=\sigma
t_{1}\sigma ^{-1}$, which makes the relations (\ref{braid}) become
cyclic. One has:
\begin{eqnarray}
\sigma t_i=t_{i-1}\sigma .\label{affine-rep1-cycle}
\end{eqnarray}
So that the affine Hecke (or T.L.) algebra is generated by the
generators $t_i$ and the cyclic operator $\sigma$ obeying
(\ref{affine-rep1-cycle}) and does not require a representation of
the permutations. $\sigma^n$ is a central element which can be set
equal to one, and $\sigma^*=\sigma^{-1}$ if we take $t^*=t^{-1}$.

Given the Hecke algebra, there is a simple realization of the affine
Hecke algebra which consists in taking $y_1=1$. Then, $\sigma$ is
defined as:
\begin{eqnarray}
\sigma=  t_{n-1}^{-1}...t_{1}^{-1} .\label{cycle}
\end{eqnarray}

Using the braid relations, one sees that $\sigma t_i=t_{i-1}\sigma$
for $ i>1$, and one can define  $t_n$  by $t_n=\sigma
t_{1}\sigma^{-1} $. Using the braid relations again, one gets
$\sigma t_n=t_{n-1}\sigma $. This defines an operator $\sigma $
which allows to construct the affine generators with
(\ref{affine-rep1}).

\subsection{Polynomial representations.\label{Polynomial
representations of the Hecke algebra}}

We consider a space of polynomials in $n$ variables $z_i,$
constructed as a linear combinations of monomials:
$z^{\mu}=z_1^{\mu_1}z_2^{\mu_2}...z_n^{\mu_n} $ with a total degree
$|\mu|=\sum \mu_i$ fixed. There is a natural action of the
permutations and of the operators $s_i$ on this space defined by:
\begin{eqnarray}
\bar\psi(z_1,..z_i..z_j..,z_n)k_{ij}&=&
\bar\psi(z_1,..z_j..z_i..,z_n),\cr
\bar\psi(z_1,..,z_i..,z_n)s_{i}&=& c
\bar\psi(z_1..,sz_i..,z_n).\label{polyaction}
\end{eqnarray}
It is convenient to consider the polynomials in an infinite set of
variables $z_i,\ i\in \cal Z,$ with the identification: $z_{i+n}=s
z_i$. The operator $\bar \sigma$ (\ref{sigma-k}) takes the form:
\begin{eqnarray}
\bar\psi\bar \sigma(z_i)=c\bar\psi(z_{i+1}).\label{sigmacycle1}
\end{eqnarray}
 The condition $\sigma^n=1$ imposes the relation $c^n
s^{|\mu|}=1$.

 As shown in (\ref{poly-hecke}), it is straightforward to derive the
following representation of the Hecke relations
(\ref{braid},\ref{hecke}):
\begin{eqnarray}
\bar t_{ij}=-q^{-1}+(1-k_{ij}){qz_i-q^{-1}z_j\over
z_i-z_j}.\label{polyactionTL}
\end{eqnarray}
In this way we obtain a representation of the affine Hecke algebra
acting on homogenous polynomials of a given total degree.

The operators $x_{ij}$ take the form:
\begin{eqnarray}
x_{ij}=-q^{-1}+(q-q^{-1})(1-k_{ij}){z_j\over
z_i-z_j}.\label{polyactionX}
\end{eqnarray}

In the appendix \ref{diag-yi}, we show that there is a natural order
on the monomial basis, $z^{\mu}$, for which the operators $x_{ij}$,
and hence the $y_i$ are realized as lower triangular matrices.

The operator $y=y_1+...+y_n$ can be seen to commute with the Hecke
generators. It is therefore equal to a constant in an irreducible
representation. Its eigenvalue evaluated on the highest weight
polynomial $P_{\lambda}$ thus characterizes the representation. It
is given by:
\begin{eqnarray}
y_{\lambda}=c(-q)^{1-n}
(s^{\lambda_1}+s^{\lambda_2}q^2+...+s^{\lambda_n}q^{2(n-1)}).
\label{valpropY}
\end{eqnarray}

 If $\lambda'$ is a permutation of the partition $\lambda$, we say that
$z^{\lambda'}$ is of degree $\lambda$. In this paper, we are mainly
concerned with the monomials $z^{\lambda'}$, of degree:
\begin{eqnarray}
\lambda=({n\over 2}-1,{n\over 2}-1,{n\over 2}-2,{n\over
2}-2,...,0,0), \label{partition}
\end{eqnarray}
and of total degree $|\lambda|={n\over 2}({n\over 2}-1)$.

We will consider the subclass $\lambda_{\pi}$ of permutations of
$\lambda$ which are smaller than $\lambda$ for the order introduced
in the appendix \ref{diag-yi}. According to the analysis made there,
$\lambda_\pi$ can be obtained from $\lambda$ by a sequence of
permutations $(\lambda'_i,\lambda'_{i+1})\to
(\lambda'_{i+1},\lambda'_{i})$ with $\lambda'_i>\lambda'_{i+1}$. The
only monomials which can be obtained that way are indexed by the
standard Young tableaus with two columns of ${n\over 2}$ boxes:
\begin{eqnarray}
 z^{\lambda_{\pi}}=(z_{\mu_1}z_{\nu_1})^{n\over
2}(z_{\mu_2}z_{\nu_2})^{{n\over 2}-1}...(z_{\mu_{n\over
2}}z_{\nu_{n\over 2}})^0,\label{youngtableau}
\end{eqnarray}
with $\mu_1>\mu_2>...>\mu_{n\over 2} $, $\nu_1>\nu_2>...>\nu_{n\over
2} $, and $\mu_i>\nu_i$. To simplify notations, we denote these
monomials by $z^{\pi}$ instead of $z^{\lambda_\pi}$.

We identify the standard Young tableaus with the paths $\pi=[h_i]$
introduced in the appendix \ref{wordrepresentation}: $h_0=h_{n}=0$,
$h_i\ge 0$ and $h_{i+1}-h_{i}=\pm 1$. These paths are obtained using
the rule: $h_i-h_{i-1}=1$ if $i\in \{\mu_j\}$, and $h_i-h_{i-1}=-1$
if $i\in \{\nu_j\}$. For the paths, we use the order $\pi\ge\pi'$,
if $[h_i]\ge[h'_i]\ \forall i,$ which coincides with the reverse
order for the monomials: $z^{\pi}\le z^{\pi'}$.

This identification is illustrated in figure
\ref{fig:wordrepresentation}.

\section{Representation of the affine T.L. algebra on words.\label{T.L. algebra}}

For $n$ even, there is a simple representation $(\mathcal{H}_n)$ of
the T.L. algebra $\mathcal{ A}_n$ obtained as follows. One considers
the left action of $\mathcal{ A}_n$ on the space $\mathcal{
A}_n\alpha$ where $\alpha$ is the minimal projector $\alpha
=e_{1}e_{3}...e_{n-1}$. A basis of this space is given by reduced
monomial words in the $e_i$. The elements of this basis can be put
into correspondence with paths or link patterns. In the appendix
\ref{wordrepresentation} we exhibit a basis of reduced words and we
define an order relation on the reduced words.

 A scalar product can be defined as:
\begin{eqnarray} \pi^* {\pi'}=\langle \pi|\pi'\rangle \alpha,
\label{produitscalaire}
\end{eqnarray}
where $e^*_i=e_i$ and the involution reverses the order of the
letters. In the link-pattern representation, this scalar product is
given by: $\tau^l$ where $l$ is the number of loops one gets by
stacking the link patterns of $\pi$ and $\pi'$ on top of each other.
If $\tau = -(q+{q^{-1}})$ with $q$ not a root of one, this scalar
product is positively definite \cite{Jones}. For this scalar product
the T.L generators $e_i$ are by construction hermitian.

To obtain the affine algebra representation, let us define as in
(\ref{cycle}) the cyclic operator:
\begin{eqnarray}
\sigma=-q^{{n\over 2}-2} t^{-1}_{n-1}...t^{-1}_{1}, \label{cycle1}
\end{eqnarray}
where the normalization is such that in the link-pattern
representation, $\sigma $ acts by cyclicly permuting the indices
$i\to i-1$ (see appendix \ref{actionofsigmaonwords}). One can define
an additional generator, $e_{n1}=\sigma e_{12} \sigma^{-1}$, which
acts in the same way as $e_{ii+1}$ with the two indices $1,n$. The
affine generators are constructed using (\ref{affine-rep1}) with
$y_1=-q^{{n\over 2}-2}$.

In the appendix \ref{triangularity}, we show that the operators
$y_i$ are realized as triangular matrices in $\mathcal{H}_n$, they
are hermitian for the choice $q=q^*$. Their sum $y=\sum_i y_i$ is
constant with a value given by:
\begin{eqnarray}
y=-(q+{q^{-1}}){q^{n\over 2}-q^{-{n\over 2}}\over q-q^{-1}}.
\label{valpropy}
\end{eqnarray}

There is an imbedding of $\mathcal{H}_{n-2}$ into $\mathcal{H}_{n}$
given by $\pi \to \pi e_{1}$ and a projection $E$ from
$\mathcal{H}_{n}$ to $\mathcal{H}_{n-2}$ given by:
\begin{eqnarray}
e_1\pi=\tau E(\pi)e_1 .\label{project-2}
\end{eqnarray}
This projection is both triangular and hermitian.

In \ref{identify}, we identify $\mathcal{H}_n$ with $\mathcal{
A}_{n\over 2}$. This allows us to interpret the projection $E$ as a
conditional expectation value of $\mathcal{ A}_{n\over
2}\to\mathcal{ A}_{{n\over 2}-1}$ \cite{Jones}.

\bigskip
\section{q-deformed Quantum Hall Effect wave functions.\label{Q.H.E.deformation}}

\subsection{Statement of the Problem.}

Let us consider a vector $\Psi$:
\begin{eqnarray}
\Psi=\sum_{\pi}\pi F_{\pi}(z_i), \label{PDF}
\end{eqnarray}
constructed in the following way. The vectors $\pi$ are the basis
vectors of $\mathcal{H}_n$ on which the  T.L. algebra acts to the
left. $F_{\pi}$ are homogeneous polynomials in the variables
$z_1,z_2,...,z_n$ ($n$ is even). The polynomial coefficients of
$\Psi$ carry a representation of the affine Hecke algebra generated
by the operators $\bar t_i$ and $\bar \sigma $ acting to the right.
The problem is to determine the coefficients $F_\pi$ in such a way
that both actions give the same result on the vector $\Psi$:
\begin{eqnarray}
\Psi\bar t_{i}&=& t_{i}\Psi \cr\Psi \bar\sigma&=&
\sigma\Psi,\label{duality}
\end{eqnarray}
The first of these relations is equivalent to the more familiar
relation (\ref{Y=k}) derived in \ref{poly-hecke}.

Said differently, we look for a dual action of the affine T.L.
algebra acting on polynomials. Unless we specify it, we address this
problem for a generic value of the parameter $q$, not a root of
unity, for which the T.L. algebra is semisimple \cite{Jones}.

\subsection{ Module $\mathcal{M}_n$.}

The dual representation of $\mathcal{H}_n$ is obtained by acting
with the T.L. generators on the dual $F_\omega$ of the highest
vector $\omega \in \mathcal{H}_n$. $\omega$ is given by the sequence
$(a_{2p+1}=p+1)$ in the characterization of words we use in the
appendix \ref{wordrepresentation} and is fully characterized by the
property that it can be written $\omega=e_i \pi$ only for $i={n\over
2}$. The dual vector $F_{\omega}$ must therefore be annihilated by
all the $e_i$ with $i\ne n/2$. We realize the module $\mathcal{H}_n$
upon acting on $F_{\omega}$ with the generators $e_i$ for $1\le i\le
n-1$. We define:
\begin{eqnarray}
\mathcal{M}_n=\mathbf{ Vec}
\{\bar\psi=F_{\omega}\psi\},\label{moduleH'}
\end{eqnarray}
where we denote with a bar $\bar \psi$ the result of the action of
the monomial $\psi$ on $F_{\omega}$. Thus we have $\bar
1=F_{\omega}$. In the appendix \ref{formedehecke}, we show that
$\mathcal{M}_n$ defined in this way is a module over the T.L.
algebra as long as the $e_i$ obey the Hecke relations
(\ref{hecke1}). In other words, the projectors
$U^-_{i,i+1}=e_ie_{i+1}e_i-e_i$ are null in $\mathcal{M}_n$. This
formal module is however not isomorphic to $\mathcal{H}_n$ unless
$F_{\omega}$ obeys some supplementary condition (\ref{fock}). Here,
we construct a representation of the T.L. algebra by identifying a
state $F_{\omega}$ dual to $\omega$ and satisfying the condition
(\ref{fock}).

\smallskip

The expression of the T.L. generators $e_i=t_i-q$ for $1\le i\le
n-1$  follows from (\ref{polyactionTL}):
\begin{eqnarray}
e_i&=&-{qz_{i+1}-q^{-1}z_{i}\over z_{i+1}-z_{i}}(1+k_{ii+1})\cr
e_i-\tau&=&(1-k_{ii+1}){qz_{i}-q^{-1}z_{i+1}\over z_i-z_{i+1}}.
\label{polynomeTLei}
\end{eqnarray}
The effect of  $e_i$ and $\tau-e_i$ is to split a polynomial
$\bar\psi$ into two polynomials belonging to $\mathcal{M}_n$, $\bar
\psi=S_1+(qz_i-q^{-1}z_{i+1})S_2$, where both $S_1$ and $S_2$ are
symmetrical under the exchange of $z_i$ and $z_{i+1}$. This
decomposition is unique and characterizes the projector $e_i$.

It can be convenient to distinguish the representation on
$\mathcal{H}_n$ from its dual on $\mathcal{M}_n$. When this is the
case, we denote $\bar e_i$ the dual projectors which act on
polynomials.
\smallskip

 One  verifies that:
\begin{eqnarray}
\Delta_{m}^{q}(z_1,...,z_p)=\prod_{1\le i<j \le m} (q z_i-q^{-1}
z_j) \label{deltadef}
\end{eqnarray}
is annihilated by all the $e_i,\ 1\le i\le m-1$, and this defines
$\Delta_{m}$ up to a  product by a symmetric polynomial in
$z_1,...,z_m.$ Therefore, the minimal degree polynomial candidate
for $F_{\omega}$ is:
\begin{eqnarray}
F_{\omega}=\Delta_{n\over 2}^q(z_1,...,z_{n\over 2}) \Delta_{n\over
2}^q(z_{{n\over 2}+1},...,z_n).\label{Fomegabar}
\end{eqnarray}
This polynomial cannot be q-antisymmetrized over  ${n\over 2}+1$
variables, since the result would have a degree at least ${n\over
2}$ in $z_1$, and this is the content of the condition (\ref{fock}).
Thus, $\mathcal{M}_n$ is a simple module which can be identified
with $\mathcal{H}_n$. This representation is characterized by its
Young diagram $(2^{n\over 2})$ having two columns of length ${n\over
2}$. Its dimension is given by the Catalan number $C_n= {n\choose
{n\over 2}}-{n\choose {n\over 2}-1}$.

\smallskip

If we denote $\pi|_{\omega}$ the coefficient of $\omega$ in the
reduced expression of $\pi$, we can identify the polynomials
$\bar\psi\in \mathcal{M}_n$ with the dual of $\mathcal{H}_n$ through
the relation: $\bar\psi(\pi)=\psi\pi|_\omega$.

We can also introduce the dual basis $F_\pi$ defined by its action
on reduced words:
\begin{eqnarray}
F_\pi(\pi')=\delta_{\pi,\pi'}.\label{vraiedualite}
\end{eqnarray}

Let $\pi_{\psi}$ be the complementary word of $\psi$ (defined in
appendix \ref{module}) such that one can write
$\psi\pi_{\psi}=\omega$ without reducing the expression. One has
$\psi\pi_{\psi}|_{\omega}=1$ and $\psi \pi|_{\omega}=0$ if
$\pi<\pi_{\psi}$ . Expanding $\bar\psi$ on the basis $F_{\pi}$, we
get: $\bar\psi=\sum_{\pi\ge \pi_\psi} \psi\pi|_{\omega}F_{\pi}$, and
by inverting the triangular system, we can obtain the expression of
$F_\pi$.

\smallskip

 Let us verify that the highest monomial of $\bar\psi$, and thus of
$F_{\pi_\psi}$ as well, is proportional to $z^{\pi_{\psi}}$. We show
this by recursion. It is true for $\psi=1$: $\bar 1=F_{\omega}$ ,
$\pi_1=\omega$ and $z^{\omega}$ is the highest monomial of
$F_{\omega}$. We assume that the property is true for $\psi'<\psi$.
If $\psi\ne1$, one can write $\bar\psi=\bar\psi' e_i$ with
$\bar\psi'<\bar\psi$, and we have $\pi_{\psi'}=e_i \pi_{\psi}$ with
$\pi_{\psi}<\pi_{\psi'}$.

Then,  according to the recursion hypothesis, the highest monomial
of $\bar \psi'$ is $z^{\pi_{\psi'}}$ which  contains the factor
$z_i^{m}z_{i+1}^n$ with $n>m$. Since $z_i^{m}z_{i+1}^n \bar
e_i=-qz_i^{n}z_{i+1}^m+ {\rm lower\ monomials},$ the highest
monomial of $\bar\psi=\bar\psi'\bar e_i$ is $z^{\pi_{\psi}}.$

We also obtain the normalization coefficient of $z^\pi$ up to a
global factor: $F_\pi=c_\pi z^\pi+{\rm lower \ monomials},$ with
$c_\pi= (-{1\over q})^{l_{\pi}}$, and $l_{\pi}$ is the number of
letters $e_i$ entering the reduced expression of $ \pi$.

\bigskip
\subsection{ Module $\mathcal{M}_n'$.\label{moduleM'}}

We now consider a larger module $\mathcal{M}_n' \supset
\mathcal{M}_n$ by letting the operator $\bar\sigma$ defined in
(\ref{sigmacycle1}) act on the polynomials. We will put some
constraint on the parameter $s$ (which characterizes $\bar\sigma$)
to have $\mathcal{M}_n'=\mathcal{M}_n$. We consider the simple case
$n=4$ in the appendix \ref{H4} and we obtain $s=q^6$ which is the
general case as we show here.

Let us assume that $\mathcal{M}_n'=\mathcal{M}_n$ and see what
constraints $s$ must satisfy to identify $\bar\sigma$ defined by its
action on polynomials (\ref{sigmacycle1}) with $\sigma$ defined in
terms of generators (\ref{cycle}).

 We observe that $\sigma^{-1}\omega_n=e_1\omega_{n-2}$, where
$\omega_{n-2}$ is the highest state in $\mathcal{H}_{n-2}$. This can
easily be verified in the link pattern representation. Thus, we must
have:
\begin{eqnarray}
\sigma E(\pi)e_1|_{\omega_n}=
E(\pi)e_1|_{\sigma^{-1}\omega_n}=E(\pi)|_{\omega_{n-2}}.\label{chaineduale}
\end{eqnarray}

 Let us consider the dual to the projection $E$, $E'$ from $\mathcal{M}_n\to \mathcal{M}_{n-2}$
defined as $\bar \psi e_1=\tau E'(\bar \psi).$ $E'$ needs to satisfy
the conditions:
\begin{eqnarray}
 &a)& \ \ E'( \bar\psi e_1)=\tau E'(\bar \psi)\cr
 &b)& \ \ E'( \bar\psi e_i)=E'( \bar\psi)e_i\ \forall i>2\cr
 &c)&\ \ E'(\bar\psi e_1)=0\Rightarrow \bar\psi e_1=0.
 \label{condition1E'}
\end{eqnarray}
>From (\ref{chaineduale}), in order to identify $\bar \sigma$ with
$\sigma$, we see that the projection $E'$ must satisfy:
\begin{eqnarray}
E'(F_{\omega_n}\bar\sigma)=F_{\omega_{n-2}}.\label{condition2E'}
\end{eqnarray}
$E'$ can be realized  as:
\begin{eqnarray}
E'(\bar \psi)=c'{1\over \phi(z,z_i)}\bar
\psi(z_1={z},z_2=q^2z,z_i),\label{definitE'}
\end{eqnarray}
where $\phi(z,z_i)=\prod_{i=3}^{n}(z_i-q^4z)$ and $c'$ is a
normalization constant. $E'$ verifies (\ref{condition1E'}a,b) by
construction as can be seen from the expression (\ref{polynomeTLei})
of $e_1-\tau$.

Using the explicit expression (\ref{sigmacycle1}) of $\sigma$, we
have:
\begin{eqnarray}
E'(F_{\omega_n}\sigma)=c' s^{-{n\over 4}-{1\over 2}}{1\over
\phi(z,z_i)}\prod_3^{{n\over 2}+1}(q^3z-q^{-1}z_i)\prod_{{n\over
2}+2}^n(q z_i-q^{-1}s
z)F_{\omega_{n-2}}(z_3,...,z_n).\label{definitE1'}
\end{eqnarray}
which imposes $s=q^6$ for the polynomial in the numerator to be
proportional to $\phi(z,z_i)$ and (\ref{condition2E'}) to be
satisfied.

\bigskip

To identify $\mathcal{M}_n$ and $\mathcal{M}_n'$, we give a more
convenient characterization of $\mathcal{M}_n'$. Consider the space
$\mathcal{M}_n''$ of homogenous polynomials in $n$ variables, and of
the minimal total degree, obeying the property:
\begin{eqnarray}
 {\rm(P):}\ \ \bar\psi(z_i=z,z_j=q^2z,z_k=q^4z)=0,\ \ {\rm if}\ i,j,k,\ {\rm are\ cyclically\ ordered}.
 \label{propiete}
\end{eqnarray}
 This property is obviously compatible with the cyclic
identification $z_{i+n}=q^6z_i$, it is thus preserved by
$\bar\sigma$ (\ref{sigmacycle1}). By applying (P) to the triplets
$(1,2,j)$, we see that the projection (\ref{definitE'}) is well
defined from $\mathcal{M}_n''$ to $\mathcal{M}_{n-2}''$.

\bigskip

We show that $\mathcal{M}_n''=\mathcal{M}_n$. For this, we first
show that $\mathcal{M}_n''$ is a module over the T.L. algebra
$\mathcal{A}_n$ and that it contains $\mathcal{M}_n$, then we show
that $\mathcal{M}_n''$ is irreducible over $\mathcal{A}_n$.

\smallskip
 To show that $\mathcal{M}_n''$ is a module over $\mathcal{A}_n$, we
verify that the generators $e_i$ preserve the property (P). Assuming
that the polynomial $\bar \psi$ verifies (P) we verify that
$\bar\psi e_i$ obeys (P) for a cyclically ordered triplet $k,l,m$.
If $\{i,i+1\}\cap \{k,l,m\}=\emptyset $, it is obvious. If $i+1=k$,
it results from the fact that $\bar\psi$ obeys (P) for the triplets
$i,l,m$ and $i+1,l,m$. The same type of argument applies if $i=m$.
If $\{i,i+1\}\subset \{k,l,m\} $, $\bar\psi(e_i-\tau)$ is
proportional to $(qz_i-q^{-1}z_{i+1})$ and therefore obeys (P).

\smallskip

Let us show that (\ref{condition1E'}c) is satisfied in
$\mathcal{M}_n''$. If $E'(\bar\psi e_1)=0$, $\bar\psi e_1$ vanishes
when $z_2=q^2z_1$, and from the definition (\ref{polynomeTLei}) of
$e_1$, it  is symmetric in $z_1,z_2$. It is therefore divisible by
$(z_1-q^2z_2)(z_2-q^2z_1)$. Hence, $\bar\psi e_1/(z_1-q^2z_2)$
satisfies (P) and has a total degree reduced by one. It is thus
equal to zero according to our minimal degree hypothesis.

\smallskip

It is clear that $F_{\omega}$ satisfies the property (P). To show
that $\mathcal{M}_n\subset \mathcal{M}_n''$, we need to show that
the degree of the polynomials in $\mathcal{M}_n''$ is the degree
${n\over 2}({n\over 2}-1)$ of $F_\omega$. We proceed by recursion on
$n$ and for the moment, we exclude the case where $e_1$ is
represented as zero in $\mathcal{M}_n''$. Due to
(\ref{condition1E'}c) there are polynomials $\bar\psi$ in
$\mathcal{M}_n''$ such that $E'(\bar\psi)\ne0$. This implies that
$\bar\psi$ has a degree at least $n-2$ in $z_1,z_2$. We can apply
the recursion hypothesis to $E'(\bar\psi)\in \mathcal{M}_{n-2}''$ to
conclude that the minimal degree is ${n\over 2}({n\over
2}-1)$.\footnote{The same argument shows that the maximal degree of
the polynomials in $\mathcal{M}_n''$ is  $\ge \lambda$ for the order
defined in \ref{diag-yi}.}

\smallskip

To show that $\mathcal{M}_n''$ is irreducible as a T.L. module, we
use the recursion hypothesis that
$\mathcal{M}_{n-2}''=\mathcal{M}_{n-2}$. Due to
(\ref{condition1E'}c), $E'$ is injective from $\mathcal{M}_n''e_1$
to $E'(\mathcal{M}_n'')\subset \mathcal{M}_{n-2}$. Since
$\mathcal{M}_ne_1=\mathcal{M}_{n-2}\subset \mathcal{M}_n''e_1$, we
have $\mathcal{M}_n''e_1=\mathcal{M}_ne_1$. Thus, if
$\mathcal{M}_n''$ contains an irreducible submodule $R\ne
\mathcal{M}_n$, $Re_1=0$. If $Re_1=0,$ from (\ref{T.L.}) we see that
all the $e_i$ are represented as $0$ in $R$, and therefore, the
polynomials in $R$ are proportional to $\Delta_{n}$ defined in
(\ref{deltadef}) times a symmetric polynomial. Since the total
degree of $\Delta_n$ is larger than ${n\over 2}({n\over 2}-1)$,
$R=0$. We conclude that $\mathcal{M}_n''=\mathcal{M}_n$ as a T.L.
module.

\smallskip

Finally, to identify $\mathcal{M}_n''$ and $\mathcal{M}_n$ as affine
modules, we observe that $y_1=\sigma^{-1}\bar\sigma$ commutes with
$\mathcal{A}_{n-1}$ generated by $e_2,...,e_n$. Since
$\mathcal{M}_n$ is irreducible over $\mathcal{A}_{n-1}$
\cite{Jones}, $y_1$ is proportional to the identity, thus $\sigma$
and $\bar\sigma$ can be identified.

\bigskip
\subsubsection{ Relation with the Macdonald Polynomials and the work of Di Francesco and Zinn-Justin.}

 As a check of consistency, we must verify that the two expressions
of the eigenvalue of the central operators $y$
(\ref{valpropY},\ref{valpropy}) are the same when $s=q^6$. This is
indeed the case if we substitute in  (\ref{valpropY}) the degree
$\lambda$ (\ref{partition}) of the highest polynomial in
$\mathcal{M}_n$ and $c=q^{3(1-{n\over 2})}$.

For a generic $s$, the operator $y$ (\ref{valpropY}) can be
diagonalized on the basis of symmetric polynomials and its
eigenvectors define the Macdonald polynomials \cite{Macdo1}. We have
seen that when $s=q^6$, the polynomial representation is reducible.
As a counterpart, some diagonal elements  $y_{\lambda'}$ of $y$
become degenerate with $y_\lambda$, for example,
$\lambda'_2=\lambda_2-1$, $\lambda'_5=\lambda_5+1$. Thus, $y$ cannot
be diagonalized. We must use another operator such as ${dy\over ds}$
to define the analogous symmetric polynomial.

In the non semisimple case $q^2+q+1=0,\ (\tau=1),$ the T.L.
representation admits a sub-representation given by ${\rm Vec}
\{\sum x_\pi \pi, {\rm\ with} \ \sum x_\pi=0\}.$ The trivial
representation $\Omega$ is obtained by equating to zero these
vectors. The dual polynomial $F_{\Omega}=\sum_{\pi}F_\pi$ is
therefore symmetrical of degree $\lambda$, and obeys the property
(P)(\ref{propiete}). This completely determines it to be
proportional to the Schur function $s_{\lambda}$ with $\lambda$
given by (\ref{partition}). Indeed, $s_{\lambda}$ has a degree
$\lambda$ and satisfies (P) since three columns of the determinant
which defines it become linearly dependant when we make the
substitution (P). By the same argument as used in \ref{moduleM'},
the degree of a symmetric polynomial satisfying (P) must be at least
$\lambda$ (relatively to the order of partitions which follows from
\ref{diag-yi}), and this proves its uniqueness.

In this limit, the $F_\pi$ are also the components the ground state
of the $O(n=1)$ transfer matrix, and thus, we have proved that this
sum is $s_{\lambda}$, which is the result of \cite{Pdf}.

It would be interesting to see if in this limit, $F_{\Omega}$ can be
recovered as the eigenvector of some operator such as ${dy\over
ds}$.

\subsection{ Representation on Gaudin's determinants.}

It is well known that the Bethe scalar products \cite{Gaudin} can be
expressed using a quotient of two determinants. Here, we construct a
representation of the T.L. algebra acting on these quotients. We
split the variables $z_i$ into $A=\{z_1,...,z_{n\over 2}\}$ and
$B=\{z_{{n\over 2}+1},...,z_{n}\}$. We also introduce $p$ a square
root of $q$, $p^2=q$. We define the polynomial $F'_{\omega}$:

\begin{eqnarray}
F'_{\omega}(z_1,..,z_n)={{\left|
(p^2z_i-p^{-2}z_j)^{-1}(pz_i-p^{-1}z_j)^{-1}\right|}\over {\left|
(pz_i-p^{-1}z_j)^{-1}\right|}}\Delta_{n}^q(z_1,...,z_n),\ {\rm\
with\ } i\in A,\ j\in B.
 \label{Gaudindeterminant}
\end{eqnarray}

The first factor is the ratio of the Gaudin determinant with the
Cauchy determinant \cite{Gaudin}. It is also related to the domain
wall boundary condition partition function \cite{korepin} of a six
vertex model with weights: $a=qx-q^{-1}y,b=px-p^{-1}y,c=\sqrt
{xy}(p-p^{-1})$ \footnote{Notice that for this six vertex model,
${a^2+b^2-c^2\over ab}=p+{ p^{-1}}\ne\tau$.}.

The second factor $\Delta_n$ (\ref{deltadef}) transforms the scalar
product into a polynomial. This factor has an innocuous effect on
the T.L. algebra since:

\begin{eqnarray}
\Delta_{n}^q(z_1,...,z_n)t_i=\tilde t_i\Delta_{n}^q(z_1,...,z_n),
 \label{innocuous}
\end{eqnarray}
where $\tilde t_i$ is obtained from $t_i$ (\ref{polyactionTL}) by
the substitution $q\to -1/q$ which preserves the relations
(\ref{braid},\ref{hecke}), but exchanges $e_i$ with $e_i-\tau$.

The ratio of the two determinants being symmetrical in the two sets
of variables $A$ and $B$, $F'_{\omega}$ is annihilated by all the
$e_i$ with $i\ne{n\over 2}$.

\smallskip

To show that the action of the T.L. algebra (\ref{polynomeTLei}) on
$F'_\omega$ produces an irreducible module, we proceed as in
\ref{moduleM'}. Consider the space $\mathcal{M}_n$ of homogenous
polynomials in $n$ variables, and of the minimal total degree,
obeying the property:
\begin{eqnarray}
 {\rm(P'):}&&\ \ \bar\psi(z_{i_1}=q^{a_1}z,z_{i_2}=q^{a_2}z,z_{i_3}=q^{a_3}z)=0,
 \ \ {\rm if}\ i_1,i_2,i_3,\ {\rm are\ cyclically\
 ordered,}\cr && {\rm and\ for:}\ \
 (a_1,a_2,a_3)=(-1,0,1),(-1,1,0),(-2,0,2),(0,-1,1).
 \label{propiete2}
\end{eqnarray}
Note that these triplets are stable under the cyclic permutation,
$(a_1,a_2,a_3)\to(a_3-2,a_2+1,a_1+1)$, and the transpositions,
$(a_i,a_{i+1})\to (a_{i+1},a_{i}),$ whenever $|a_{i+1}-a_{i}|=1$.

>From the cyclic invariance, we deduce that this space is preserved
under the action of $\sigma$ (\ref{sigmacycle1}) if we take $s=q^3$.

By applying  the property (P') to $z_1,z_2,z_i$ with
$(a_1,a_2,a_3)=(-1,1,0)$ and $(-2,0,2)$, we can define a projection
(\ref{definitE'}) from $\mathcal{M}_n\to \mathcal{M}_{n-2}$. The
polynomial $\phi(z,z_i)$ is now a product of two factors
$\phi(z,z_i)=\prod_{i=3}^{n}(qz-z_i)(q^4z-z_i)$. Arguing as in
\ref{moduleM'}, we see that this projection satisfies the properties
(\ref{condition1E'}).

This space is stable under the action of the generators $e_i$. The
proof is similar to the one given in \ref{moduleM'}  and requires
the stability of the triplets $(a_1,a_2,a_3)$ under the
transpositions. The minimal degree is now $n({n\over
2}-1)=2|\lambda|$ with $|\lambda|$ given by (\ref{partition}).

\smallskip

Let us show that $F'_{\omega}$ (\ref{Gaudindeterminant}) satisfies
the property (P') (\ref{propiete2}). We consider $(i_1,i_2,i_3)$ and
$(a_1,a_2,a_3)$. If the variables $z_l,z_m$  with $l<m$,
corresponding to two $a_i$ which differ by $2$, belong to the same
set $A$ or $B$, $F'_\omega(z_m=q^2z_l)=0$ due to the factor
$\Delta_n$. Otherwise, two variables $z_l=z\in A$ and $z_m=q^2z\in
B$ differ by a factor $q^2$. By isolating the contribution of the
pole $(p^2z_l-p^{-2}z_m)$ in the Gaudin determinant, we factorize a
term $\prod_i(qz-z_i)$ coming from the Cauchy denominator, and this
enables to conclude that
$F'_\omega(z_{i_1}=q^{a_1}z,z_{i_2}=q^{a_2}z,z_{i_3}=q^{a_3}z)=0$ in
all the other cases.

\smallskip

Arguing as in \ref{moduleM'} we conclude that $\mathcal{M}_n$ is an
irreducible module over the affine T.L. algebra and that it
coincides with the module obtained upon acting with the generators
on $F'_\omega$.

\smallskip

We verify again that the eigenvalue of the central operators $y$
(\ref{valpropY}) is given by (\ref{valpropy}). Now, $s=q^3$ instead
of $q^6$ in \ref{moduleM'}, but the degree $2\lambda$
(\ref{partition}) of the highest polynomial in $\mathcal{M}_n$ is
doubled and $c$ keeps the same value $c=q^{3(1-{n\over 2})}$.

In the nonsemisimple case $q^2+q+1=0$, using the result of
\cite{stroganov} we have $F'_\pi= s_{\lambda}F_\pi$, and therefore,
$\sum_\pi F'_{\pi}= s_{\lambda}^2$.

\section{Conclusion.}

Let us conclude with a few comments and questions.

On the mathematical side, this work provides a unification ground
around the conjectures relating the eigenvector components of a loop
model transfer matrix, the six vertex model domain wall boundary
condition partition function and other mathematical objects. It
opens the possibility to deform the polynomials underlying these
conjectures by presenting them from the algebra representation point
of view. We believe that these conjectures are related to
incompressibility, and we hope to return to this point in a future
publication.

>From a technical point of view, it would be interesting to repeat
the Jones construction of \ref{identify} on the polynomials
directly. This would allow to recover in a direct way the product
structure which they carry since they are associated to elements of
the T.L. algebra.

The precise correspondence between the  polynomial obtained here
and the Macdonald polynomials needs to be clarified.

Finally, do the deformed wave functions considered here have
anything to do with physics? At this moment, we have no answer to
this question. A step towards a physical interpretation would be to
identify a scalar product and a Hermitian Hamiltonian for which
these wave functions are the ground states. This could also be
useful to access to the excited states which play an important role
in the Quantum Hall Effect.

\subsection{Acknowledgements}

I wish to thank Philippe di Francesco for generously explaining me
his works and for discussions.

I am greatly indebted to Kirone Mallick, Gregoire Misguich and
particularly Bertrand Duplantier for their help during the course of
this work.

\bigskip

\appendix
\section{Word representation.\label{wordrepresentation}}

\subsection{ Reduced words.}

The module $\mathcal{H}_n$ is obtained by acting with the T.L.
generators of $\mathcal {A}_n$ on the lowest state
$\alpha=e_1e_3...e_{n-1}$. Using the relations (\ref{T.L.}), we
obtain a basis of $\mathcal{H}_n$ given by reduced words $\pi$:
\begin{eqnarray}
\pi=(e_{a_{n-1}}e_{a_{n-1}+1}..e_{n-1})...(e_{a_{2p+1}}e_{a_{2p+1}+1}..e_{2p+1})...
(e_{a_3}e_{a_3+1}..e_3)e_1,\label{word}
\end{eqnarray}
with, $a_{2p+1}\le 2p+1$, and $1<a_3<..< a_{2p+1}<..<a_{n-1} $. So,
a word is fully characterized by the sequence $(a_{2p+1})$.

On reduced words there is a natural order relation: $\pi>\pi'$ if
$\pi$ is written $b{\pi'}$ with b a monomial. One has $\pi\ge\pi'$
if $a_{2p+1}\le a'_{2p+1}$ for all $p$.
\smallskip

Another way to represent a reduced word is in terms of paths. Let
$m_i$ be the number of times the generator $e_i$ appears in the
reduced expression of $\pi$. One has $m_{2i}=m_{2i-1} {\rm \ or}\
m_{2i-1}-1$ and $m_{2i+1}=m_{2i} {\rm \ or}\ m_{2i}+1$. We define
$h_{2i}=2m_{2i}-1$, $h_{2i-1}=2m_{2i-1}-2$ and $h_0=h_n=0$ by
convention. We can describe the words $\pi$ by the paths $\pi=[h_i]$
where $h_0=h_{n}=0$, $h_i\ge 0$ and $h_{i+1}-h_{i}=\pm 1$. Using the
path representation, one has $\pi\ge\pi'$, if $[h_i]\ge[h'_i]\
\forall i.$

If $\pi'$ is a non reduced word, by reducing it, one decreases the
number of times the generator $e_i$ appears in its expression. We
thus see that the order relation can be presented in a weaker form
valid for non reduced words: If $\pi'$ is a word, not necessarily
reduced and $\pi$ is a reduced word, $\pi>\pi'$ if $\pi'$ can be
obtained by erasing letters  $e_k$ from the (reduced) expression of
$\pi$.

\smallskip
Finally, there is way to characterize this representation in terms
of link patterns. It is convenient to dispose the $n$ points
cyclically around a circle. A link pattern is obtained by pairing
all the points in the set $\{1,2,...,n\}$:
$\pi=\{[i_1,i_2],[i_3,i_4],...,[i_{n-1},i_n]\}$, in such a way that
two links never cross. In practise, if $[i,j]$ is a link, then the
other links $[k,l]$ are either inside, or outside the interval
$[i,j]$. The action of $e_{i,i+1}$ is given by:
$e_{i,i+1}\{[i,i+1],...,[i_{n-1},i_n]\}=\tau
\{[i,i+1],...,[i_{n-1},i_n]\}$, and
$e_{i,i+1}\{[i,j],[i+1,k]...,[i_{n-1},i_n]\}=\{[i,i+1],[j,k],...\}$.
In this representation, $\alpha=\{[1,2],[3,4],...,[n-1,n]\}$, and
$\omega=\{[1,n],[2,n-1],...,[{n\over 2}-1,{n\over 2}+1]\}$.

\begin{figure}

\begin{center}

\includegraphics[width=6cm]{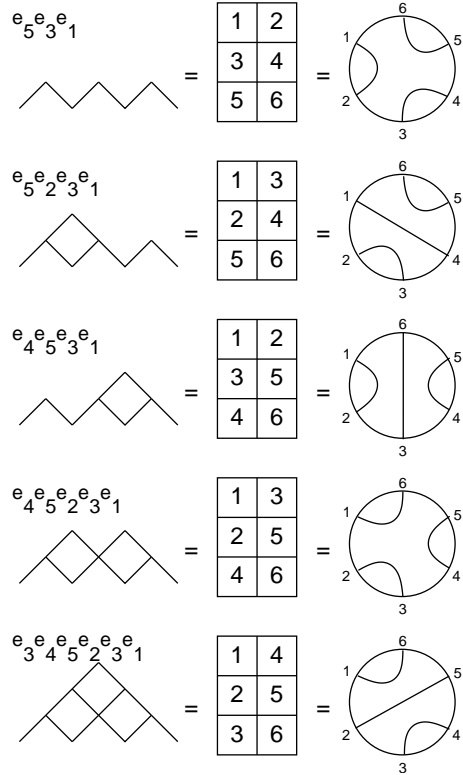}

\caption[99]{

The three different ways to represent a word illustrated in the case
of $\mathcal{H}_6$.}

\label{fig:wordrepresentation}

\end{center}

\end{figure}

These representations are illustrated in figure
\ref{fig:wordrepresentation}.

\subsection{ Identifying $\mathcal{H}_n$ with $\mathcal{A}_{n\over
2}.$\label{identify}}

The link pattern representation allows to identify in a natural way
$\mathcal{H}_n$ with $\mathcal {A}_{n\over 2}.$ If we split
$\{1,2,...,n\}$ into two subsets :$ \{1,2,...,{n\over 2}\}$ and $
\{{n\over 2},...,n\}$, the product $\pi*\pi'$ is defined on the link
patterns by ``stacking'' the two link patterns and identifying the
last ${n\over 2}$ points of $\pi$ with the first ${n\over 2}$ points
of $\pi'$ through $i\equiv n+1-i$. The link pattern $\pi*\pi'$ is
obtained by removing the loops which appear in this concatenating
operation by giving them a weight $\tau$.

Another identification can be achieved on paths by folding a path of
length $n$ into a loop of length ${n\over 2}$. In this way, we
realize $\mathcal {A}_{n\over 2}$ as the algebra of double paths
acting on Bratteli diagrams
\cite{Pasbratel}\cite{Pasthese}\cite{Jones}.

In this identification, $\mathcal {A}_{n\over 2}$ is a bimodule over
itself. The first ${n\over 2}-1$ generators $e_i\in \mathcal
{A}_{n}$ are identified with the generators of $\mathcal {A}_{n\over
2}$ acting to the left, while the last ${n\over 2}-1$ generators are
identified with $e_{n+1-i}$ acting to the right.

The state $\omega$ is the identity in $\mathcal {A}_{n\over 2}$, and
the trace in $\mathcal {A}_{n\over 2}$ coincides with the scalar
product with $\omega$ in $\mathcal {A}_{n}$:
\begin{eqnarray}
{\rm t r} (x)=\tau^{-{n\over 2}}\langle \omega |x\rangle .
\label{identifytrace}
\end{eqnarray}

The projection: $E_{{n\over 2}}=\sigma^{-{n\over
2}+1}E\sigma^{{n\over 2}-1},$ with $E$ given by (\ref{project-2})
can be reinterpreted as a conditional expectation value
\cite{Jones}, $E_{n\over 2}:\ \mathcal {A}_{n\over 2}\to \mathcal
{A}_{{n\over 2}-1}$. Jones construction enables then to construct
$e_{n\over 2}\in \mathcal {A}_{{n\over 2}+1}$ algebraically from the
knowledge of  $E_{n\over 2}$.

\bigskip \bigskip
\subsection{ Triangularity of $y_m$.\label{triangularity}}

Let us show that the affine generators
$y_{m+1}=t^{-1}_mt^{-1}_{m-1}...t_1^{-2}...t^{-1}_m$ are triangular
in the word representation. It is obvious for $y_1=1$ and
$y_2=t_1^{-2}$ since $e_1$ is triangular. We proceed by recursion
and assume that $y_{k}$ are triangular for $k<m+1$. Using these
hypotheses, we show that $y_{m+1}$ is also triangular.

First we show that $y_{m+1}$ acts diagonally on $\alpha$. To study
the action of $y_{m+1}$ on $\alpha$, we distinguish the two cases
$m$ odd or even. If m is odd, then:
\begin{eqnarray}
y_{m+1}\alpha=t^{-1}_my_{m}t^{-1}_me_m...=-qt^{-1}_my_{m}e_m...=-\lambda_{m}qt^{-1}_me_m...=q^2\lambda_{m}\alpha,
\end{eqnarray}
where $\lambda_m$ is the eigenvalue of $y_{m}$ on $\alpha$. If $m$
is even, we make use of the fact that
$t_{m-1}^{-1}t^{-1}_me_{m-1}={1\over q}e_me_{m-1}$ and the same
relation with the indices $m$ and $m-1$ exchanged to obtain:
\begin{eqnarray}
y_{m+1}\alpha&=&t^{-1}_mt^{-1}_{m-1}y_{m-1}t^{-1}_{m-1}t^{-1}_me_m...={1\over
q}t^{-1}_my_{m-1}e_me_{m-1}...\cr &=&{1\over
q}\lambda_{m-1}t^{-1}_mt^{-1}_{m-1}e_me_{m-1}...={1\over
q^2}\lambda_{m-1}\alpha.
\end{eqnarray}
We deduce that $\alpha$ is an eigenstate of $y_m$ with the
eigenvalue $\lambda_{m}$ obeying the recursion relations
$\lambda_{2m}=q^2\lambda_{2m-1}$, $\lambda_{2m+1}={1\over
q^2}\lambda_{2m-1}$. Together with the fact that $\lambda_1=1$, we
deduce (\ref{valpropY}).

\bigskip

To show that $y_{m+1}$ is triangular on words $\ne \alpha$. We
proceed by recursion and assume that $y_{m+1}$ acts in a triangular
way on words  $<\pi$ and show that the property is also true for
$\pi$.

Let us consider the action of $y_{m+1}$ on a reduced word $\pi\ne
\alpha$. This word can be put under  the form $\pi=e_i\pi'$ where
$\pi'<\pi$. We consider the three cases, $i\ne m,m+1$, $i=m$,
$i=m+1$. In the third case, either the word can be written in the
form $e_{m+1}e_m\pi'$ with $\pi'$ reduced, or it can be written
$e_p\pi'$ with $p<m$. The second possibility reduces to the first
case and we need only consider the first possibility.

We observe that $y_{m+1}$ commutes with $e_i$:
$y_{m+1}e_i=e_iy_{m+1}$ if $i>m+1$ or if $i<m$. It is obvious if
$i>m+1$ and follows from the braid relations if $i<m$. In the three
cases we can thus write:

\begin{eqnarray}
y_{m+1}e_i\pi'&=&e_i(y_{m+1}\pi')\ \ {\rm for}\ i\ne m,m+1,\cr
y_{m+1}e_m\pi'&=&t^{-1}_my_{m}t^{-1}_{m}e_m\pi'=-qt^{-1}_m(y_{m}e_m\pi'),\cr
y_{m+1}e_{m+1}e_m\pi'&=&t^{-1}_my_{m}t^{-1}_{m}e_{m+1}e_m\pi'=t^{-1}_m(y_{m}e_m\pi'
 +{1\over q}y_me_{m+1}e_m\pi').
\end{eqnarray}
It follows from the hypothesis that the terms in brackets are less
than $\pi$. In the first case because $\pi'<\pi$, and in the two
others because $y_m$ is assumed to be triangular.

To conclude that $y_{m+1}$ is triangular, we must show that the
action of $e_i$ in the first case and  $e_m$ in the two other cases
preserves the triangularity : If $e_i\pi$ is a reduced word and
$\pi'\leq \pi$, then, $e_i\pi'\leq e_i\pi$. If $e_m\pi$ is a reduced
word and $\pi'\leq e_m\pi$, then $e_m\pi'\leq e_m\pi$. Finally, if
$e_{m+1}e_m\pi$ is a reduced word and $\pi'\leq e_{m+1}e_m\pi$, then
$e_m\pi'\leq e_{m+1}e_m\pi$. These properties follow from the weak
form of the order relation. This concludes the proof of
triangularity of $y_{m+1}$.

\bigskip
\subsection{  Action of $\sigma$ on words.\label{actionofsigmaonwords}}

The action of $\sigma=-q^{{n\over 2}-2}t_{n-1}^{-1}...t_1^{-1}$ on
words can be computed similarly. First, using the braid relation
(\ref{braid}), one sees that $\sigma e_i=e_{i-1}\sigma$ for $i>1$.
To fully characterize its action, we must compute $(\sigma \alpha)$.
Using $t_1^{-1}e_1=-qe_1$ and $t_{m+1}^{-1}t^{-1}_me_{m+1}={1\over
q}e_me_{m+1}$  we obtain:
\begin{eqnarray}
\sigma \alpha= \prod_{i=1}^{{n\over 2}-1}e_{2i}\alpha.
\end{eqnarray}
Thus, $(\sigma \alpha)$ can be characterized by the property:
\begin{eqnarray}
e_{2i}(\sigma \alpha)=\tau (\sigma \alpha),
\end{eqnarray}
for $1\le i\le {n\over 2}$.  $(\sigma\alpha)$ can then be used as a
lowest state to construct a canonical basis by acting on it with the
generators $e_2,...,e_{n}$.

\bigskip

\section{Explicit construction of $\mathcal{M}_4$.\label{H4}}

Let us construct $\mathcal{M}_4$ the dual of $\mathcal{H}_4$. The
basis of $\mathcal{H}_4$ is given by the words $e_1e_3=\alpha,\
e_2e_1e_3=\omega$. So we search for a vector $\Psi$ of the form:
\begin{eqnarray}
\Psi=F_{\alpha}(z_1,..,z_4)\alpha+F_{\omega}(z_1,..,z_4)\omega,\label{Psi4}
\end{eqnarray}
where $F_{\alpha},F_{\omega}$ are polynomials of degree $(1,1)$ in
the variables $z_i$.
 The action of the $T.L.$ affine algebra is given by the matrices:
\begin{eqnarray}
e_1=e_3=\pmatrix{\tau&1\cr 0&0}, \ e_2=e_4=\pmatrix{0&0\cr 1&\tau},\
\sigma =\pmatrix{0&1\cr 1&0}. \label{TL4}
\end{eqnarray}

We can obtain the dual representation by acting with the generators
on  $F_{\omega}\equiv \pmatrix{0,1}$ annihilated by $\bar
e_1=e_1,\bar e_3=e_3$. The minimum degree polynomial annihilated by
$ e_1, e_3$ is given by:
\begin{eqnarray}
F_{\omega}=(qz_1-q^{-1}z_{2})(qz_3-q^{-1}z_{4}). \label{F2-4}
\end{eqnarray}

Let us take $\bar\sigma=\sigma$ of the form:
\begin{eqnarray}
F(z_1,z_2,z_3,z_4)\sigma =c F(z_2,z_3,z_4,sz_1). \label{C-4}
\end{eqnarray}
We obtain two different expression for $F_{\alpha}\equiv(1,0)$ which
we must equate. One  results from the cyclic property:
$F_{\alpha}=F_{\omega}\sigma$, the other given by:
$F_{\alpha}=F_{\omega}(e_2-\tau)$.

We get the equation:
\begin{eqnarray}
{(qz_1-q^{-1}z_2)(qz_3-q^{-1}z_4)-(qz_1-q^{-1}z_3)(qz_2-q^{-1}z_4)\over
z_2-z_3}=c (qz_4-q^{-1}sz_1), \label{miracle-4}
\end{eqnarray}
which determines $s=q^6$, $c=q^{-3}$, and:

\begin{eqnarray}
F_{\alpha}=(qz_2-q^{-1}z_{3})(q^{-2}z_4-q^{2}z_{1}). \label{F2-4}
\end{eqnarray}

\bigskip

\section{ Module $F_{\omega}$.\label{module}} Let us define a T.L.
module $M$ defined in terms of a state $F_{\omega}$ satisfying
$F_{\omega }e_i=0$ for $i\ne {n\over 2}$. The module is obtained by
acting with the T.L. generators and reducing words using the T.L.
relations (\ref{T.L.}). In this module, a canonical basis is:
\begin{eqnarray}
\bar \psi=F_{\omega}(e_{n\over 2}e_{{n\over 2}-1}...e_{a_{n\over
2}+1}e_{a_{n\over2}})...
(e_{p}e_{{p}-1}...e_{a_{p}})...(e_{n-1}...e_{a_{n-1}}),\label{word1}
\end{eqnarray}
where the $p$ take the all the values between ${n\over 2}$ and $n-1$
and the $a_{p}$ are restricted by the conditions: $a_{p}\le p+1$,
$a_{n\over 2}<a_{{n\over 2}+1}..< a_{p}<..<a_{n-1}$. The convention
is that if $a_p=p+1$, the sequence $(e_{p}...e_{a_{p}})$ is empty. A
word $\bar \psi$ is fully characterized by the sequence $(a_{p})$.
The word can also be associated to the Young diagram
$[\mu_{p+1-{n\over 2}}]=[p-a_p+1]$.

There is a reflection symmetry, $i\to n-i$, and an alternative
description of the module in terms of reflected words:
\begin{eqnarray}
\bar \psi=F_{\omega}(e_{n\over 2}...e_{b_{n\over 2}-1}e_{b_{n\over
2}})...(e_{p}...e_{b_{p}-1}e_{b_{p}})...(e_{1}...e_{b_{1}}),\label{word11}
\end{eqnarray}
$1\le p\le {n\over 2}$, $b_p\ge p-1$, $b_{n\over 2}>...>b_1$. It is
associated to the dual Young diagram $[\mu'_{{n\over
2}-p+1}]=[b_p-p+1]$

A similar order relation as defined earlier holds for reduced words,
$\bar\psi'<\bar\psi$ if $\bar\psi$ can be written
$\bar\psi=\bar\psi'a$. For non reduced words $\bar\psi'$, it is
sufficient that $\bar\psi'$ can be obtained by erasing letters $e_k$
from the (reduced) expression of $\bar\psi$.

\smallskip
 In general, the module $F_{\omega}$ is reducible, it will
be irreducible if $F_{\omega}$ satisfies the Fock condition:
\begin{eqnarray}
F_{\omega}(1+\sum_{m=0}^{{n\over 2}-1} q^{m+1}t_{{n\over
2}}...t_{{n\over 2}-m})=0.\label{fock}
\end{eqnarray}

In this case, The only allowed words $\bar\psi$ (\ref{word1}) can be
associated to their complementary $\pi_{\psi}$ in such a way that
one can write without reducing the expression:
\begin{eqnarray}
\psi\pi_{\psi} =\omega.\label{motcomplement}
\end{eqnarray}
Thus, we get the supplementary constraint $a_{p}>2p+1-{n}, \
b_p<2p-1$.

\subsection{ Reducing the Hecke Module to its T.L. form.\label{formedehecke}}

Let us consider a module $M'$ over the Hecke algebra defined by
acting with the Hecke algebra generators satisfying (\ref{hecke1})
on the state $F_{\omega}$ satisfying $F_{\omega} e_i=0$ for $i\ne
{n\over 2}$ . We want to show that the Hecke algebra acts as a T.L.
algebra on this module. For this, we first show that the Hecke
relations (\ref{hecke1}) are sufficient to reduce the word basis of
$M'$ to be of the T.L. form (\ref{word1}). Thus, $M'$ and $M$ can be
identified as vector spaces. From this, we will deduce that $M'=M$
as modules. In other words, the projectors
$U^-_{i,i+1}=e_ie_{i+1}e_i-e_i$ are null in $M'$.

\smallskip
Let us assume that it is not true. Since all the basis elements of
$M'$ are obtained upon acting on $F_{\omega}$ with letters $e_k$,
there is a basis element $\bar\psi e_i$ which cannot be expressed as
a linear combination of words of the form (\ref{word1}) although
$\bar\psi$ is of the form (\ref{word1}). Among all the $\bar\psi$
which verify this property, we can take the smallest possible for
the order relation, so that that $\bar\psi'e_i$ is of the form
(\ref{word1}) when $\bar\psi'<\bar\psi$. We show that this leads to
a contradiction.

Let us consider the word $\bar\psi e_i$. It is a word of the form
(\ref{word1}) in the three following cases. When $\bar\psi e_i$ is a
reduced word $>\bar\psi$, for $i=a_p-1$ if $a_p-1>a_{p-1}$. When
$\bar\psi e_i=\tau \bar\psi$ when $i=a_p$ and $a_p>a_{p+1}-1$. When
$\bar\psi e_i=0$ if $i<a_{n\over 2}-1$ or $i>b_{n\over 2}+1$.

The two remaining cases to consider are: First, when $a_p< i<
a_{p+1}-1$ for some $p$. Second, when $a_p<i \le a_p+k$ if
$a_{p+k}=a_p+k$ with $k\ge 1$. The second case can be studied
similarly to the first one using the reflection symmetry $i\to n-i$
and corresponds to $b_{p'}>i>b_{p'-1}+1$.

In the first case, $\bar\psi
e_i=\bar\psi'(e_p...e_{a_p+1}e_{a_p})e_i(e_{p+1}...e_{a_{p+1}})...$,
and using the relation (\ref{hecke1}), we see that:
\begin{eqnarray}
e_p...e_{a_p+1}e_{a_p}e_i      = e_{i-1}e_p ...e_{a_p+1}e_{a_p}+
e_p... e_{i+1}(e_i-e_{i-1}) e_{i-2}...e_{a_p+1}e_{a_p}.\label{sueur}
\end{eqnarray}
The second term is $<\bar\psi$ and therefore of the T.L. form by the
recursion hypothesis. The first term can be eliminated by repeating
this relation $p-{n\over 2}$ times to push $e_{i}$ and then
$e_{i-1},...,e_{i+{n\over 2}-p}$ to the left of the word. The last
application of the relation gives a term $F_{\omega} e_{i+{n\over
2}-p-1}=0$ since $i+{n\over 2}-p-1<{n\over 2}$.

This exhaust all the possibilities and $\bar\psi e_i$ can always be
expressed as a linear combination of reduced T.L. words
(\ref{word1}) in contradiction with the hypothesis. Therefore, the
word basis of $M'$ coincides with the word basis (\ref{word1}).

\smallskip
To conclude that $M'=M$, let us consider the projectors
$U^-_{i,i+1}=e_ie_{i+1}e_i-e_i$, and the space $M''\subset M'$
annihilated by all the $U^-_{i,i+1}$. The space $M''$ defines a
module for the T.L. algebra. Since $F_\omega\in M''$, this module
can be identified with $M$. Therefore, $M$ is a subspace of $M'$
with the same dimension, and thus, $M=M'$.

\bigskip

\section{Yang-Baxter Equation and Polynomials:}

\subsection{Polynomial representation of the Hecke generators
\label{poly-hecke}}

In this section, we derive the expression of the Hecke generators
$\bar t_i$ (\ref{polyactionTL}) from the Yang-Baxter equation.

The Yang-Baxter algebra \cite{Gaudin} (also called $RLL=LLR$
relation) can be expressed as:
\begin{eqnarray}
 R_{12}(z_1,z_2)L_1(z_1)L_2(z_2)=L_2(z_2)L_1(z_1)R_{12}(z_1,z_2),
\label{yang-baxter1} \end{eqnarray} where $R_{12}(z_1,z_2)$ is a
solution of the Yang-Baxter equation:
\begin{eqnarray}
 R_{12}(z_1,z_2)R_{13}(z_1,z_3)R_{23}(z_2,z_3)=R_{23}(z_2,z_3)R_{13}(z_1,z_3)R_{12}(z_1,z_2).
\label{yang-baxter0} \end{eqnarray}

If we assume that $R_{12}(z_1,z_2)=Y_{12}(z_1,z_2)P_{12}$ where
$P_{12}$ acts in the natural way on the spin indices,
$(P_{12}t_{13}=t_{23}P_{12}),$  but commutes with $z_i$,
(\ref{yang-baxter1}) rewrites as:
\begin{eqnarray}
 Y_{12}(z_1,z_2)L_2(z_1)L_1(z_2)=L_2(z_2)L_1(z_1)Y_{12}(z_1,z_2)=L_2(z_1)L_1(z_2)k_{12},
\label{yang-baxter2}
\end{eqnarray}
where $k_{12}$ acts to the left by permuting the variables
$z_1,z_2$. If we normalize of $Y(z_1,z_2)$ so that:
\begin{eqnarray}
Y_{12}(z_1,z_2)Y_{12}(z_2,z_1)=1, \label{YY-yang-baxter2}
\end{eqnarray}
it is consistent to demand that the $Y_{ii+1}$ act as a
representation of the permutation algebra on some wave function
$\Psi$:
\begin{eqnarray}
 Y_{12}(z_1,z_2)\Psi(z_1,z_2)=\Psi(z_2,z_1)=\Psi(z_1,z_2)k_{12}.
\label{Y=k}
\end{eqnarray}
The  $Y_{ij}$ are are called Yang's operators in \cite{Gaudin}.

A well known solution of
(\ref{yang-baxter0}),(\ref{YY-yang-baxter2}) in terms of the Hecke
algebra (\ref{hecke}) is:
\begin{eqnarray}
 Y_{12}(z)={ t_{12}-z t_{12}^{-1}\over z q -
 q^{-1}},
\label{yang-baxter3}
\end{eqnarray}
where $z={z_1\over z_2}$.

Substituting (\ref{yang-baxter3}) in (\ref{Y=k}), we can also
rewrite this relation as:
\begin{eqnarray}
 t_{12}\Psi(z_1,z_2)=\Psi(z_1,z_2)\bar t_{12},
\label{t=g}
\end{eqnarray}
Where $\bar t_{12}$ takes the form:
\begin{eqnarray}
\bar t_{12}=-q^{-1}+(1-k_{12}){qz_1-q^{-1}z_2\over z_1-z_2},
 \label{polyactionTL1}
\end{eqnarray}
and this coincides with (\ref{polyactionTL}).

\subsection{Commutation relations of the affine generators $y_i$.\label{discussion}}

We motivate the commutation relation (\ref{affine-gener}c) from the
Yang-Baxter algebra (\ref{yang-baxter1}) point of view. This also
reveals a complete symmetry between the spectral parameters $z_i$
and the generators $y_i$.

Let us substitute the spectral parameters $z_i$ with the affine
generators $y_i$ in $L(z_i)$, and we require that the relation
(\ref{t=g}) are preserved under the action of the algebra $L_i$ on
$\Psi$:
\begin{eqnarray}
t_{12}L_1(y_1)L_2(y_2)\Psi=L_1(y_1)L_2(y_2)\Psi\bar t_{12},
 \label{consistance}
\end{eqnarray}
assuming that (\ref{t=g}) holds for $\Psi$.

To avoid cumbersome expressions, we use from here the transposed
notation $\bar a X$ for $X \bar a$. We must then transpose back the
final algebraic relations we deduce. In the transposed notations
(\ref{consistance}) is equivalent to:
\begin{eqnarray}
 (t_{12}-\bar t_{12})L_1(y_1)L_2(y_2)=0,
 \label{consistance1}
\end{eqnarray}
under the hypothesis that $t_{12}=\bar t_{12}$ to the right of any
expression. Let us for the moment assume that $\bar t_{12}$ commutes
with the symmetrical expressions in $y_1,y_2$.

After substituting the expression of $L_i(y_i)$ deduced from
(\ref{yang-baxter3}):
\begin{eqnarray}
 L_1(y_1)=(y t_{10}-y_1 t_{10}^{-1})P_{01},
 \label{expressionLi}
\end{eqnarray}
the term proportional to $y^0$ requires that $\bar t_{12}$ commutes
with $y_1 y_2$, while the term proportional to $y$ imposes that:
\begin{eqnarray}
 (t_{12}-\bar t_{12})(y_2t_{01}t_{12}^{-1}+y_1t_{01}^{-1}t_{12})=0,
 \label{consistance2}
\end{eqnarray}
under the hypothesis that $t_{01}=\bar t_{12}$ to the right of any
expression. This gives:
\begin{eqnarray}
 y_2\bar t_{12}+(q-q^{-1})y_1-\bar t_{12}y_1&=&0,\cr
 y_1-\bar t_{12}y_2\bar t_{12}&=&0,
 \label{consistance3}
\end{eqnarray}
which is equivalent to $y_2\bar t_{12}=\bar t_{12}^{-1}y_1$ and
implies in particular that $\bar t_{12}$ commutes with the
symmetrical expressions in $y_1,y_2$. After transposition, it yields
(\ref{affine-gener}c) back.

\smallskip

Alternatively, we can substitute $z_i$ for $y_i$ in
(\ref{affine-gener}c) and verify that the relation is obeyed when we
use the expression (\ref{polyactionTL1}) of $\bar t_i$.

\subsection{\bf Eigenvalues of the  $y_j$ in the polynomial
case.\label{diag-yi}}

We show that the operators $y_j$ defined with the polynomial
representation \ref{Polynomial representations of the Hecke algebra}
are triangular matrices. Let us recall the expression of $y_i$:
\begin{eqnarray}
y_i&=&x_{ii+1}x_{ii+2}...x_{in-1}s_ix_{i1}...x_{ii-1}\label{yang-rep00}
\end{eqnarray}
where the operator $x_{i,j}$ takes the form for $i<j$:
\begin{eqnarray}
x_{ij}=-q^{-1}+(q-q^{-1})(1-k_{ij}){z_j\over
z_i-z_j},\label{polyactionX}
\end{eqnarray}
and the operators $s_i$ act as:
\begin{eqnarray}
 P(z_1,..,z_i..,z_n)s_{i}&=& c
P(z_1..,sz_i,...,z_n).\label{polyaction-si}
\end{eqnarray}

$x_{12}$ commutes with $z_1z_2$ and with $z_k$ for $k\ne 1,2$. It
acts in the following way for on the monomials $z_1^m,\ z_2^m $:
\begin{eqnarray}
z_1^mx_{12}&=&-q^{-1}z_1^m+(q-q^{-1})(z_1^{m-1}z_2+z_1^{m-2}z_2^2+...+z_2^m)\cr
z_2^mx_{12}&=&-qz_2^m-(q-q^{-1})(z_1^{m-1}z_2+z_1^{m-2}z_2^2+...+z_1z_2^{m-1})
\label{ordreX}
\end{eqnarray}
Let us consider which new monomials $z^{\lambda'}$ can appear when
one acts with $x_{12}$ on the monomial $z^{\lambda}$. First, all the
$\lambda'_l$ for $l\ne 1,2$ are equal to $\lambda_l$. Then, if
$\{\lambda'_i\lambda'_j\}\ne \{\lambda_i\lambda_j\}$ with
$\{i,j\}=\{1,2\}$ and $\lambda'_j\le \lambda'_i$ , we must have
$\{\lambda'_i,\lambda'_j\}= \{\lambda_i-p,\lambda_j+p\}$ for some
integer $p$. Finally, if $\{\lambda'_1\lambda'_2\}=
\{\lambda_1\lambda_2\}$, the only possibility is that:
$(\lambda'_1,\lambda'_2)=(\lambda_2,\lambda_1)$ with
$\lambda_1>\lambda_2$.

Let us  define an order on the  monomials by saying that
$z^{\lambda}$ is larger than $z^{\lambda'}$ if either $\lambda'$ is
obtained from $\lambda$ by a sequence of squeezing operations
$\{\lambda_i,\lambda_j\}\to \{\lambda_i-1,\lambda_j+1\}$ with
${\lambda_i>\lambda_j+1},$ or $\lambda'$ is a permutation of
$\lambda$ and can be obtained from $\lambda$ by a sequence of
permutations $(\lambda_i,\lambda_{i+1})\to
(\lambda_{i+1},\lambda_{i})$ with $\lambda_i>\lambda_{i+1}$ . It
follows from the above analysis that the action of $y_j$ on a
monomial produces only monomials which are smaller with respect to
this order. Thus the eigenvalues of the operators $y_j$ are given by
the diagonal elements in the monomial basis.

Given the partition $\lambda=(\lambda_1,...,\lambda_n)$, the
eigenvalues corresponding to the monomials associated to it are  all
obtained by permutations of the multiplet:
\begin{eqnarray}
(y_j)=c(-q)^{1-n}(t^{\lambda_j}q^{2(j-1)}) .\label{spectre-yj}
\end{eqnarray}

\end{document}